\documentclass[acmtog]{acmart}
\renewcommand\footnotetextcopyrightpermission[1]{} 
\usepackage{booktabs}
\usepackage{graphicx} 
\usepackage{multirow} 
\usepackage{makecell} 
\usepackage{natbib}
\usepackage{hyperref}
\usepackage{subcaption}

\AtBeginDocument{%
  \providecommand\BibTeX{{%
    \normalfont B\kern-0.5em{\scshape i\kern-0.25em b}\kern-0.8em\TeX}}}

\setcopyright{none} 
\begin{document}

\title{U.S. Election Hardens Hate Universe}

\author{Akshay Verma}

\email{averma29@gwu.edu}
\affiliation{%
  \institution{The George Washington University}
  \country{USA}
}

\author{Richard Sear}
\affiliation{
    \institution{The George Washington University}
    \country{USA}
}

\author{Neil Johnson}
\affiliation{
    \institution{The George Washington University}
    \country{USA}
}

\begin{abstract}  
{\bf Local or national politics can trigger potentially dangerous hate in someone  \cite{schackmuth2018extremism, starbird_disinformations_2019,shahi_protecting_2023,obaidi2022great,noauthor_stress_nodate,johnson_hidden_2019,cynthia_miller-idriss_hate_2020,douek2022content}. But with a third of the world's population eligible to vote in elections in 2024 alone \cite{shahi_protecting_2023}, we lack understanding of how individual-level hate multiplies up to hate behavior at the collective global scale. Here we show, based on the most recent U.S. election, that offline events are associated with a rapid adaptation of the global online hate universe that hardens (strengthens) both its network-of-networks structure and the `flavors' of hate content that it collectively produces. Approximately 50 million potential voters in hate communities are drawn closer to each other and to the broad mainstream of approximately 2 billion others. It triggers new hate content at scale around immigration, ethnicity, and antisemitism that aligns with  conspiracy theories about Jewish-led replacement \cite{obaidi2022great} before blending in hate around gender identity/sexual orientation, and religion. Telegram acts as a key hardening agent -- yet is overlooked by U.S. Congressional hearings and new E.U. legislation \cite{euDSA}. Because the hate universe has remained robust since 2020, anti-hate messaging surrounding not only upcoming elections but also other events like the war in Gaza, should pivot to blending multiple hate ‘flavors’ \cite{lupu_offline_2023} while targeting previously untouched social media structures.}

\end{abstract}

\maketitle

\begin{figure*}
  \centering
  \includegraphics[width=\linewidth]{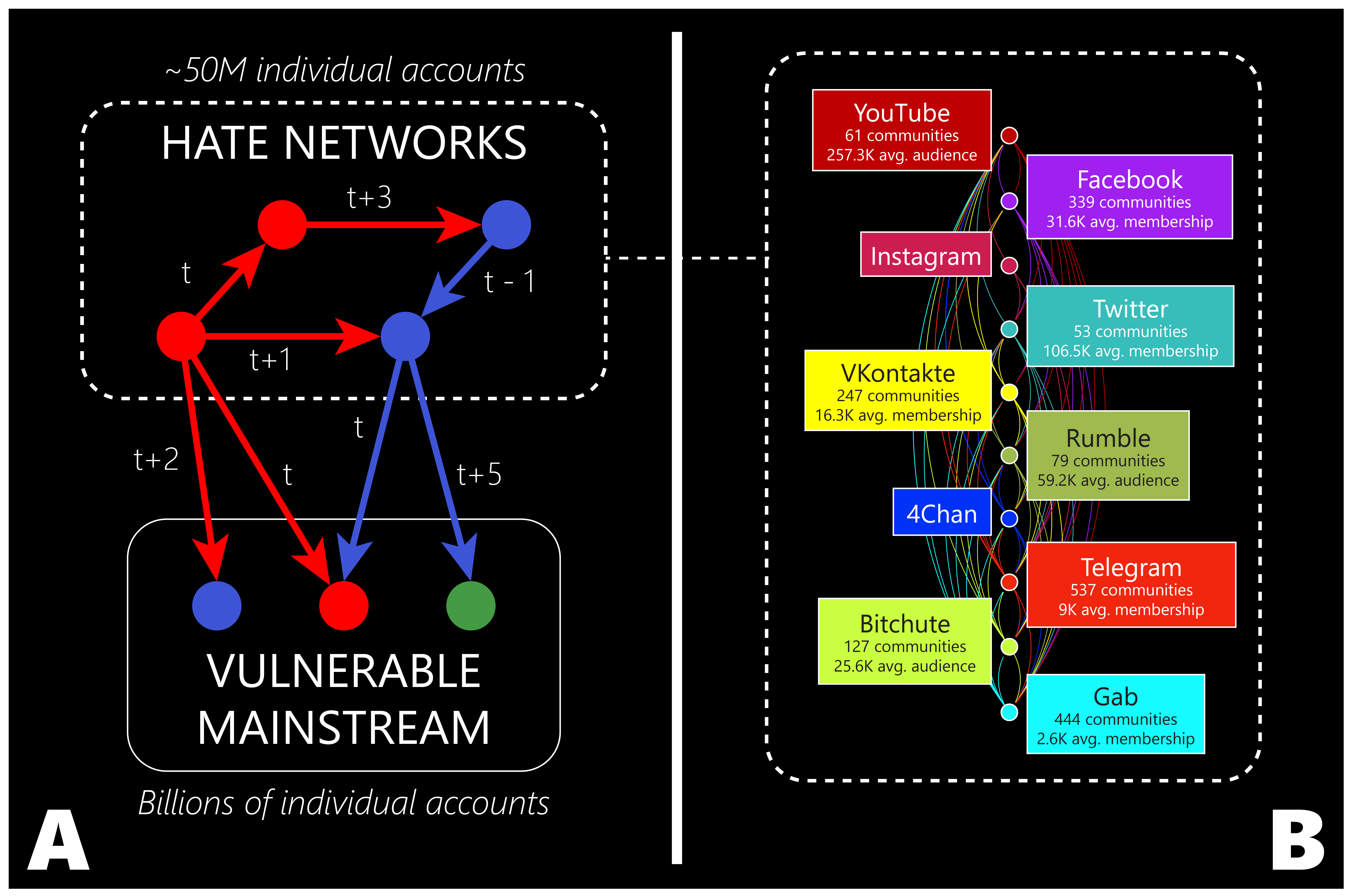}
  \caption{Hate universe. \textbf{A}: Schematic of how a hate community (node) on a given social media platform (given color) establishes a link (by sharing a URL for a piece of content) to another community (node) at a given time. A link from node A to node B means that members of community A are alerted to B's existence, and can visit community B to share hate. Each community is a platform-provided community: a Telegram Channel, a Gab Group, a YouTube Channel, etc. Top dotted box is the subset of communities that are identified as hate communities. These link together over time to form hate networks-of-networks. Bottom box represents communities that are directly linked to by hate communities but which are not themselves in the list of hate communities; hence, we label these as vulnerable mainstream. \textbf{B}: Empirically determined composition of the hate universe. Boxes show number of hate communities (nodes) in a given platform and the average number of members of each community. Edges shown are aggregated over time. \\
    \Description{Schematic dynamic structures in the online hate network, as well as an actual visualization of the connections and quantities in this network}}
    \label{fig:hu3}
\end{figure*}

The Internet is a breeding ground for hate \cite{schackmuth2018extremism, starbird_disinformations_2019}. Hate content and its supporters thrive in hate networks comprising an interconnected web of in-built  communities across multiple social media platforms \cite{johnson_hidden_2019}. Each in-built community (e.g. Telegram Channel, Gab Group, YouTube Channel) shows an ideology rooted in hatred and discrimination, and it can have  anywhere from a few to a few million members. 
Any member of such a `hate community' (community A) can at any time cross-post content from another community (community B) and hence create a link (hyperlink) from A to B (see Fig. \ref{fig:hu3}A). Other members of A are then alerted to B's existence, and can visit community B to share their hate \cite{noauthor_stress_nodate,liu_control_2016, cynthia_miller-idriss_hate_2020,johnson_hidden_2019, gelfand_cultural_2021, douek2022content}. Reference \cite{Zheng_Sear_Illari_Restrepo_Johnson_2024} provides explicit examples of links and hate communities (nodes).

A `hate universe' is hence formed by the set of all such hate communities across all platforms -- together with the other communities that they directly link to (vulnerable mainstream, Fig. \ref{fig:hu3}A).
It is a novel network-of-networks in which both the network structure and its content (hate narratives) co-evolve dynamically on similar timescales \cite{watts2004small}.
While there have been many in-depth studies on the relationship between polarizing real-world events and the evolution of online hate \cite{rachel_brown_new_2018, cinelli_dynamics_2021,chen_why_2015, chen_why_2015_1, kim_trust_2013}, our focus on the dynamics of this hate universe at scale enables us to fill a gap in current understanding of how hate evolves globally around a local or national event such as an election. We focus here on the U.S. presidential election and choose the most recent case (2020) since it is the first in which many small and large platforms feature prominently, and its offline events included the  Capitol attack on the day that Congress certified the president-elect \cite{lee2022storm}. 

We build the hate universe using the same methodology as prior publications \cite{lupu_offline_2023,Zheng_Sear_Illari_Restrepo_Johnson_2024,velasquez_online_2021} but now expanded to more platforms. We define a hate community as one for which multiple subject matter experts determine its most recent activity displays hate towards at least one of the protected classes listed in the FBI's definition of hate crimes, or extreme racial identitarianism. Prior publications contain a full discussion and concrete examples \cite{lupu_offline_2023,Zheng_Sear_Illari_Restrepo_Johnson_2024,velasquez_online_2021}. We first  extracted an initial list of hate communities (see data provided) and monitored for cross-posts (links). This then led us to new communities and hence allowed us to grow a comprehensive final list of hate communities. Repeating this process in a semi-automated way eventually produced a crude representation of the entire public-facing hate universe across social media platforms (Fig. \ref{fig:hu3}B). 
In this directed network (i.e. comprising links from one community to another), each source node is a hate community (Fig. \ref{fig:hu3}A top panel). If its target node is not on our hate community list, we label it as vulnerable mainstream: it is possible that it includes discriminatory content, but it had not at the time qualified to be on the hate community list.

We classify the text in the hate communities' posts using a trained natural language processing (NLP) model \cite{lupu_offline_2023} to identify types (`flavors') of hate. Seven were classified: race, gender, religion, antisemitism,  gender identity/sexual orientation (GISO), immigration, and ethnicity/ identitarian/nationalism (EIN) \cite{lupu_offline_2023}. This published method has been checked against human reviewers and has a high degree of accuracy \cite{lupu_offline_2023}.

Our social media data collection is, like others, necessarily imperfect. However, the resulting hate universe does contain several billion individual accounts. Given that most of their owners are likely over the voting age in their respective countries, and given that more than 2 billion people worldwide are due to vote in the near term, it provides a crude first-cut. Our focus is on  changes in this hate universe's topology and content (i.e. narratives) through the 2020 presidential election (November 3) and  subsequent Capitol attack (January 6, 2021). We are not concerned here with the extent to which these changes have a causal effect on the offline events or are simply a response.

\begin{figure*}
  \centering
  \includegraphics[width=\linewidth]{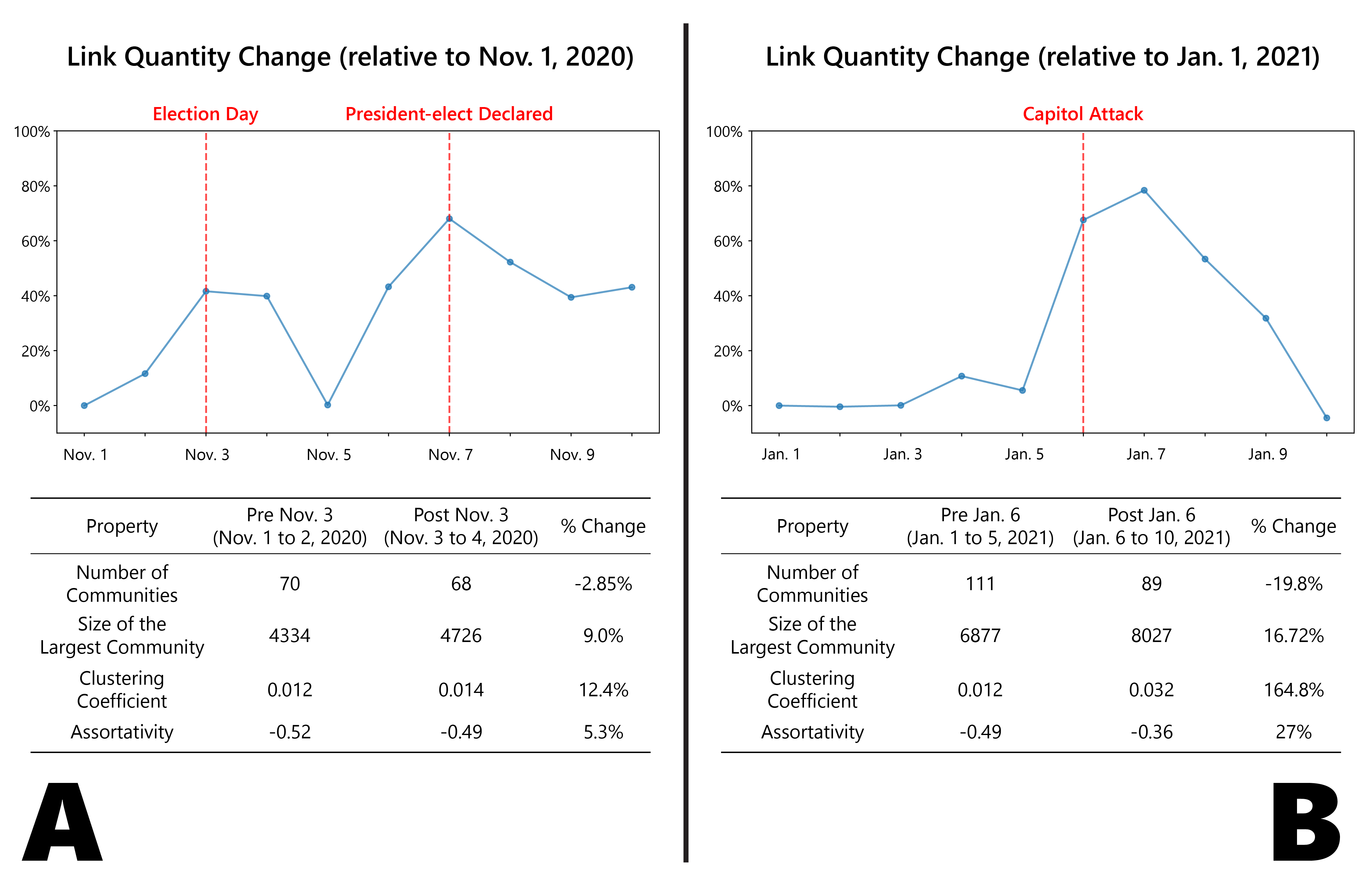}
  \caption{Changes in key network metrics  surrounding events in the 2020 U.S. election period, specifically the actual date of the election (\textbf{A}) and the day of Congress' confirmation which is when the attack took place on the U.S. Capitol (\textbf{B}) \\
    \Description{Plots and tables showing network metric changes before and after Nov. 3 (left) and Jan. 6 (right)}}
    \label{fig:network-metrics}
\end{figure*}

\section*{Findings}

Figure \ref{fig:network-metrics} shows there was a surge in the creation of hate links (i.e. links from hate communities) during both the election itself (November 2020) and the follow-up (January 2021). On November 3, the day of the election, the increase was 41.6\% compared to November 1. On November 7 when Joe Biden was declared president-elect, this number spiked even further at 68\%. An even larger spike surrounds January 6. Going forward in the digital age, the message of this for policymakers is that any event of seemingly only local or national interest has the capability to instantly trigger hate activity globally.

\subsection*{Hate Universe's Structure Hardens}
Just because links get individually added, does not automatically mean the hate universe as a whole will benefit in terms of hardening (strengthening) its structural cohesion. However, the lower panels of network measures in Fig. \ref{fig:network-metrics} show this is indeed exactly what happens. 

The clustering coefficient, which is a standard measure that captures the tendency of nodes to form connected groups of three, jumped by 164.8\% after January 6. Nodes previously on the periphery of the core formed new connections, leading to a denser network with well-defined clusters. Hence, individual nodes within the core became more interconnected on average \cite{watts2004small}. The assortativity, which is a standard measure that captures the tendency of nodes to connect with other nodes that are similar in terms of their connectivity, also increased by 27\%. This further supports the notion of increased cohesion: individual communities (nodes) preferentially connected with others who shared similar network characteristics. These suggest a strengthening of existing ideologies within the network, fostering a more homogeneous environment that is then more resilient to outside intervention \cite{leifeld2018polarization, oehlers2021graph}.

The changes in the number of communities and the size of the largest community (Fig. \ref{fig:network-metrics} lower panels) provide further evidence of  this network strengthening (hardening). The number of communities decreased by  19.8\%\, at the same time as the size of the largest community grew by 16.72\%, reflecting a convergence of smaller communities towards a single more dominant one. This evolution toward a more cohesive hate universe with a more unified ideology at scale, means it can better amplify the spread of hate speech or coordinated actions at scale. Most importantly, this is all essentially organic and hence not the result of top-down control by a single bad actor.

\begin{figure*}
  \centering
  \includegraphics[width=\linewidth]{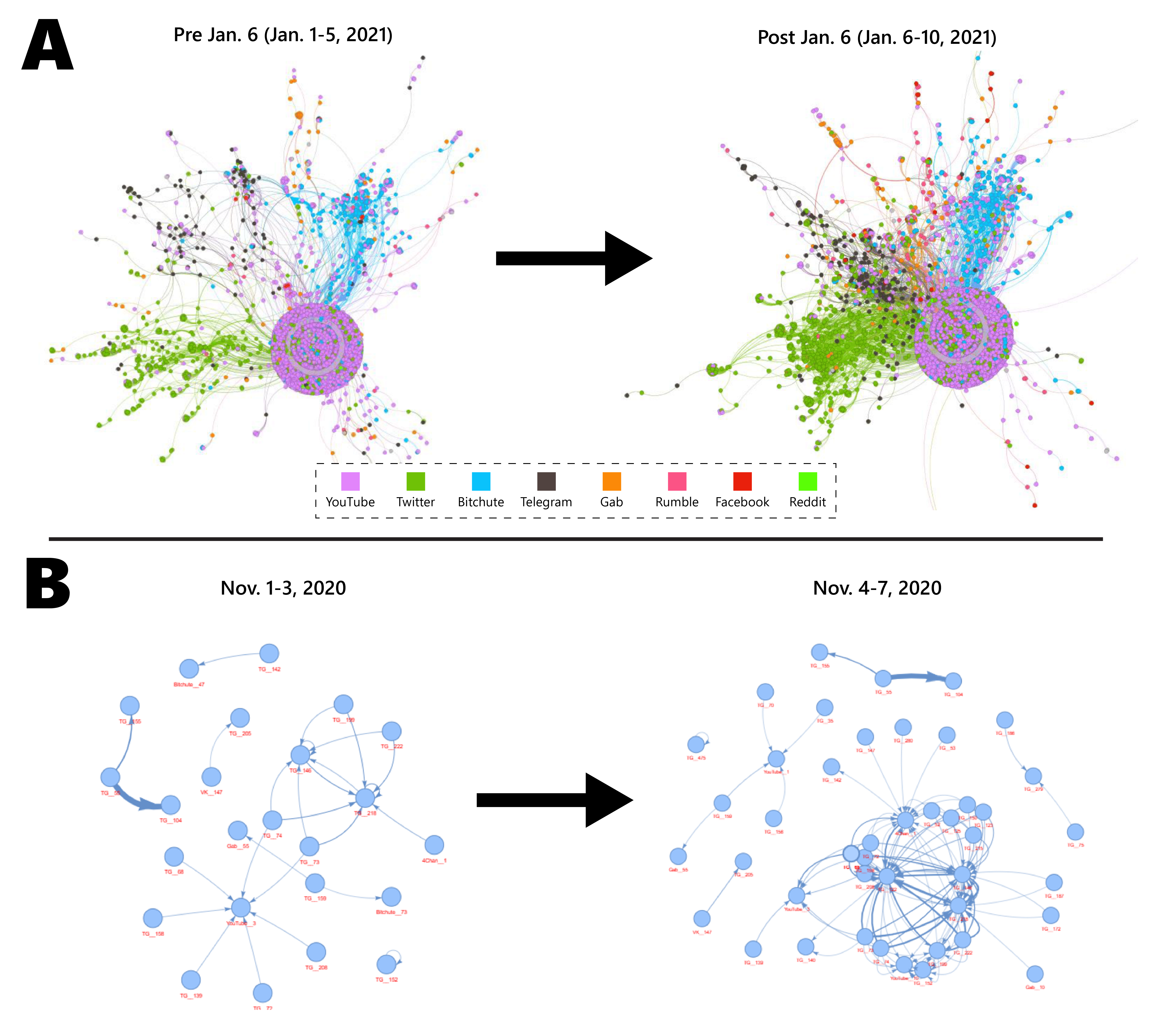}
  \caption{Hardening of hate universe's network-of-networks. \textbf{A}: Gephi visualization of the hate universe before and after January 6, 2021. The partial rings comprise dense clouds of vulnerable mainstream communities (nodes) with successively increasing numbers of links from the hate networks, hence creating orbital-like rings. \textbf{B}: Subset of  Telegram-connected networks before and after election day, show its key role as a binding agent. \\
    \Description{Social media networks colored by platform}}
    \label{fig:network-cohesion}
\end{figure*}

These conclusions are further supported by visualizations employing the ForceAtlas2 layout \cite{Jacomy_Venturini_Heymann_Bastian_2014}. This layout algorithm assigns forces to nodes, attracting or repelling them based on their edges which act like springs: this means that the network's resulting visual appearance reflects its true structure at scale. The resulting visualizations (Fig.~\ref{fig:network-cohesion}A) reveal a notable shift towards a more cohesive network-of-networks structure -- in agreement with the metrics.

\begin{figure*}
  \centering
  \includegraphics[width=\linewidth]{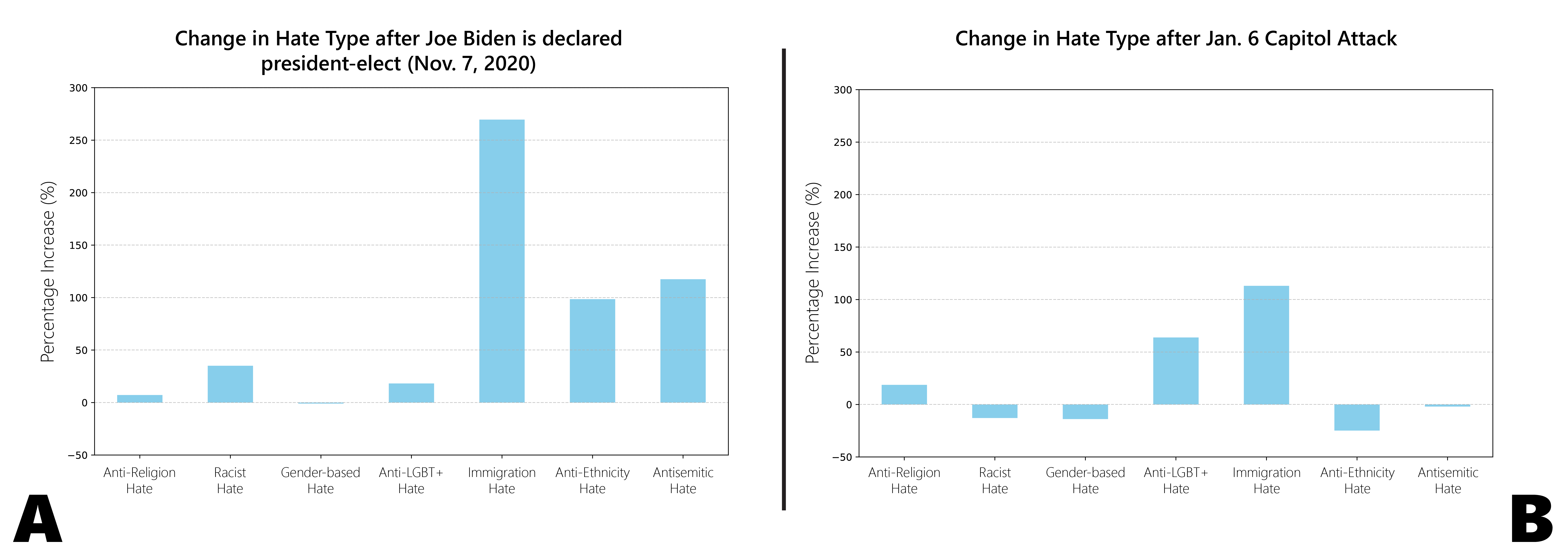}
  \caption{Hardening of hate universe's content around particular types (`flavors') of hate narratives. Results shown for the periods around the declaration of Biden as president-elect (\textbf{A}) and around January 6 (\textbf{B}). \\
    \Description{Bar graphs showing percent changes in hate type following significant events}}
    \label{fig:flavors-change}
\end{figure*}

\subsection*{Hate Universe's Narratives Harden}

Just because the hate universe's structure becomes more cohesive does not automatically mean that the type of hate content that it contains should change -- or if it does, how it would change. Yet Fig. \ref{fig:flavors-change}A shows there was a significant uptick in hate speech targeting immigration, ethnicity, and antisemitism around November 7. In particular, there was a 269.5\% surge in anti-immigration sentiments, a 98.7\% rise in ethnically-based hatred, and a 117.57\% escalation in expressions of antisemitism from November 7-11 compared to November 2-6. A comparable surge in anti-immigration sentiments occurred around January 6 (Fig. \ref{fig:flavors-change}B). There was a 108.69\% rise in anti-immigration content between January 6-10 compared to January 1-5  2020. 

These trends suggest an increase in immigration anxieties harbored by far-right communities, which often align with the Great Replacement conspiracy theory and attribution of perceived demographic shifts to Jewish influence \cite{obaidi2022great}.

We also find strong correlations between  changes in the hate universe's structure and changes in its content -- specifically, correlations between  daily link count originating from specific social media platforms and hate content over the period November 1, 2020 to January 10, 2021 (see Supplementary Figure 1). For example, there is an increase in links originating from 4Chan, Gab, Twitter, while Telegram exhibits a strong correlation with instances of hate speech that target immigration, race, and gender/sexual identity (GISO).

As well as showing how individual-level hate gets expressed collectively at scale, these findings reveal the following shortcomings in current practice and regulations:

\noindent (1) Particular platforms can play different network roles in the hate universe, in particular smaller and/or less well-known ones such as Telegram (see discussion below). This suggests that current policies that focus only on particular platforms based on their popularity (e.g. Facebook, Twitter, or TikTok) or which treat all platforms the same, will not be effective in  curbing hate and other online harms. 

\noindent (2) The type of hate content that surges around a real-world event is not necessarily the closest in theme to that event. Hence, the messaging of anti-hate campaigns ahead of an event should not focus exclusively on that event's theme. 

In summary, the hate universe's rapid self-adaptation at scale, its wide array of target-ready mainstream communities, and the  variety of hate `flavors' that it offers them, suggest that each future election will provide it with a diverse set of new recruits globally while simultaneously strengthening (hardening) its   structure and narratives at scale. These  findings suggest that anti-hate policies ahead of nominally local or national events such as elections, should mix multiple hate themes (`flavors') and use the multi-platform hate universe map to target their impact at the global scale.

\subsection*{Key Role of Telegram}

Telegram never features in U.S. Congressional hearings involving social media company representatives, nor does it feature in the E.U.'s flagship Digital Services Act that has recently come into operation. 
However, Fig. \ref{fig:network-cohesion}B reveals Telegram's key role as a `glue' in the hate network's hardening, by showing the subset of the hate universe in which Telegram is a source or target node. 
During November 4–7, there was a remarkable surge in connectivity within Telegram, as evidenced by a substantial 299\% increase in the number of connections involving Telegram compared to November 1–3 (from 592 to 2366). Telegram's role as a target node increased from 18.22\% to 33.47\%, while its role as a source node increased from 21.73\% to 37.24\%, as a percentage of all links present in the hate-to-hate sub-network. This highlights Telegram's growing significance as a central platform for communication and coordination among hate communities \cite{schulze2022far} and also aligns with growing concern among law enforcement agencies because of its association with `Q-anon' and `Pro-Trump Conspiracy theories' \cite{cnn2021,guhl2020safe,walther_telegram_2021}. 

The Telegram community we call "TG\_\_122" in our data -- which is currently named "US Voter Fraud \& Coup-Ops Intel” -- is an interesting example. It was initially absent from the network during November 1-3, but then emerged and rapidly became one of the most crucial and interconnected nodes. "TG\_\_122" exhibits strong associations with two other Telegram channels, "TG\_\_218", a right-wing podcast,  and "TG\_\_146", a right-wing Telegram channel named "Exposing Cultural Marxism", both of which were highly active before November 3.

\vskip1in

\section{Data availability}
The datasets used in this study are at \url{https://github.com/gwdonlab/data-access/tree/elections-online-hate/elections-online-hate}. The original data contain sensitive information from social media platforms: hence to comply with data protection standards and avoid potential misuse, the raw data cannot be shared publicly. However, the processed derivative datasets  can be used to reproduce the results in the study.

\section{Code availability}

Plots were created using Python (Matplotlib \cite{Hunter:2007}). The code is available at \url{https://github.com/gwdonlab/data-access}.
The network visualization in Fig. \ref{fig:network-cohesion}a was created using Gephi and the network visualization in Fig.  \ref{fig:network-cohesion}b was created using Python (PyVis).
Together, this provides readers with access to the minimum dataset that is necessary to interpret, verify, and extend the research in the article. Cosmetic changes to figures were carried out in Adobe Illustrator, a commercially available piece of software.

\section{Funding}

N.F.J. is supported by US Air Force Office of Scientific Research awards FA9550-20-1-0382 and FA9550-20-1-0383, and by The John Templeton Foundation.

\bibliographystyle{ieeetr}
\bibliography{citation}

\end{document}